\documentclass[10pt,conference]{IEEEtran}
\usepackage[left=0.6in,right=0.6in,top=0.7in,bottom=1.0in]{geometry}
\usepackage[utf8]{inputenc} 
\usepackage[T1]{fontenc}
\usepackage{url}

\usepackage{breakurl}
\usepackage{ifthen}
\usepackage{graphicx}
\usepackage[T1]{fontenc}
\usepackage[bookmarks=false]{hyperref}

\usepackage{enumitem}
\usepackage[cmex10]{amsmath}
\usepackage{amssymb}
\usepackage{multicol}
\usepackage{caption}
\usepackage{subcaption}
\usepackage{cite}
\usepackage{amsthm}
\usepackage{textcomp}
\usepackage{siunitx}
\usepackage{color}
\usepackage{cases}
\usepackage[font={small}]{caption}
\usepackage{epstopdf}
\usepackage{graphicx}
\usepackage{caption}
\usepackage{verbatim} 
\usepackage{bbm}
\usepackage{mathtools}

\newtheorem{lemma}{Lemma}

\theoremstyle{definition}

\newtheorem{definition}{Definition}

\theoremstyle{remark}

\newtheorem*{theorem*}{Theorem}

\usepackage[ruled,vlined]{algorithm2e}
\SetAlFnt{\footnotesize} 
\setlist[description]{style=multiline}

\begin{document}
\sloppy
\title{6G EdgeAI: Performance Evaluation and Analysis}

\author{
\IEEEauthorblockN{ Chien-Sheng Yang$^{*}$, Yu-Jen Ku$^{*}$, Yuan-Yao Lou$^{\dag}$, Nathan Tenny$^{*}$ and Alex C.-C. Hsu$^{*}$} 
\IEEEauthorblockA{$^{*}$ MediaTek Inc.        \quad       $^{\dag}$ Purdue University
\vspace{-2mm}
}
}
\maketitle

\begin{abstract}
Generative AI (GenAI) services powered by large language models (LLMs) increasingly deliver real-time interactions, yet existing 5G multi-access edge computing (MEC) architectures often treat communication and computing as separate domains, limiting their ability to meet stringent latency requirements. To address this challenge, we introduce an Integrated Communication and Computing (ICC) framework where computing capabilities are enabled to reside directly in radio access network (RAN) nodes and jointly manage bandwidth and computing resources. Our queueing-theoretic analysis shows that ICC outperforms 5G MEC, achieving higher service capacity (defined as the maximum arrival rate that maintains a specified fraction of jobs completed within a given delay budget) by $98\%$. We corroborate these gains through system-level simulations that account for transformer-based LLM workloads, realistic GPU specifications, and a priority-based scheduling scheme. The simulations show that ICC improves service capacity by $60\%$, demonstrating its potential to enable efficient, cost-effective real-time GenAI services in 6G.

\end{abstract}

\section{Introduction}
Generative AI (GenAI) services are rapidly expanding their reach to mainstream consumer markets, driven primarily by advances in large language models (LLM). These models enable sophisticated tasks such as real-time language translation, and content generation—capabilities with stringent low-latency requirements. As an increasing number of applications integrate GenAI functionalities, ensuring minimal response time becomes critical for delivering seamless user experiences. 

The evolution of cellular technology from 4G to 5G has dramatically expanded the types of applications that mobile networks can support, including virtual reality, augmented reality, and Industry 4.0 use cases. However, GenAI services introduce additional computational overhead, particularly to handle LLM inference workloads that existing 5G infrastructure may struggle to satisfy. Thus, 6G development aims to incorporate native computing capabilities into the mobile network fabric, supported by industry-wide efforts (e.g., AI-RAN Alliance\cite{AIRAN}).

Many of today's LLM-based services rely on large-scale cloud servers. However, this design paradigm entails significant drawbacks. The need to route requests and responses between remote data centers and end-user devices can introduce noticeable latency, limiting the potential for real-time interactions. Additionally, reliance on centralized computing inflates network congestion and raises concerns about scalability and operational overhead. Although 5G multi-access edge computing (MEC) seeks to mitigate these issues by moving compute closer to end users—thus reducing signal travel distance and improving responsiveness—current MEC implementations are typically confined to operator-managed sites located behind a User Plane Function (UPF), limiting proximity to end users. This setup rules out RAN-based or device-based servers and prevents mobile network operators (MNOs) from fully harnessing the performance, cost, and flexibility benefits of placing computing resources across the broader network, including the RAN and user devices.

Furthermore, current 5G MEC architectures typically treat communication and computing resources as separate domains, each governed by its own framework. While adequate for many existing applications, this disjoint approach cannot fully optimize the system for the stringent latency requirements of LLM inference. When both communication overhead and compute latency are critical, coordinating them independently can lead to inefficiencies and underutilized capacity. Therefore, a holistic approach that manages communication and computing resources under one unified framework is essential for 6G networks.

This paper introduces the concept of \textit{Integrated Communication and Computing} (ICC). With ICC, computing capabilities can be enabled within RAN nodes, allowing efficient offloading of LLM inference tasks  while maintaining seamless control over both communication and computing. By coordinating resource usage end-to-end, ICC offers superior performance compared to traditional disjoint approaches, opening the door for new service paradigms in which AI computations are distributed intelligently across the MNO's network.

To evaluate the benefits of ICC, we first develop a theoretical framework using queueing modeling where user requests arrive following a Poisson process and all incoming jobs contend for both communication and computing resources. Compared to 5G MEC, our queueing analysis demonstrates that integrating RAN computing with joint communication-computing latency management increases the achievable service capacity (defined as the maximum arrival rate that maintains a specified fraction of jobs completed within a given delay budget) by $98\%$.

Beyond theoretical insights, we conduct simulation-based evaluations of LLM inference tasks offloaded to an MNO's network. Our methodology accounts for typical transformer architectures, GPU specifications, and the interplay of communication and computing latencies. We further include a priority-based scheme to highlight tangible performance gains from a joint latency management approach. Simulation results validate ICC's ability to reduce end-to-end job latency and increase service capacity by $60\%$, all while lowering the operator's compute costs by $27\%$ against 5G MEC. Collectively, these findings underscore the feasibility and promise of deploying ICC for GenAI services in 6G networks.\\

\noindent\textbf{Related Work:} Recent years have witnessed a growing emphasis on integrated communication and computing (see e.g., \cite{NGA_DWC, MTK_WP, 10251878}). The 6G network is expected to facilitate ambient intelligence (AmI), revolutionizing the interaction between humans and technology. Beyond mere reactions and responses, AmI infers and anticipates user needs by collecting, processing, analyzing, and learning from a wide variety of data sources to deliver the best response in the most intuitive form possible to the end user~\cite{MTK_WP}. To realize AmI capabilities at scale, advancements in computing technologies, such as GPUs, NPUs, and DPUs, have enabled distributed inference and edge AI~\cite{9606720}, which are crucial for supporting AmI in 6G networks. Specifically, the heterogeneity of the device cloud and RAN cloud fosters tight collaboration for the distributed AI/ML inference at the edge of 6G network~\cite{10251878,10884554}. The prior studies have addressed this by partitioning inference tasks and distributing them among device clouds, edge clouds, and traditional clouds through adaptive joint device selection and LLM model partition problems~\cite{10818760} or by optimizing DNN partitioning to minimize delays~\cite{10597111}. 
On the other hand, the previous works~\cite{10759588} have focused on jointly allocating communication and computation resources to optimize LLM inference, considering heterogeneous resource constraints and user requirements. There are also existing researches that investigate the concept of device clouds built around user devices in a multi-level inference framework by utilizing open dataset~\cite{10835146} or deploying a real testbed~\cite{li2024dynamic}.

Despite these prior works, we investigate LLM inference under an ICC-based 6G framework that brings computing resources closer to the end user and jointly manages communication and computing latencies, with particular emphasis on service capacity and cost efficiency. Our queueing-based theoretical analysis shows that ICC outperforms both 5G MEC and cloud-centric models in delivering higher service capacity. We validate these insights using a system-level simulator (SLS) that integrates L1 physical layer and L2 protocol modeling for communication latency, as well as an LLM-centric inference model for computing latency under a joint latency management.

\section{Integrated Communication and Computing}
\begin{figure}[t] 
    \centering \includegraphics[width=1\linewidth]{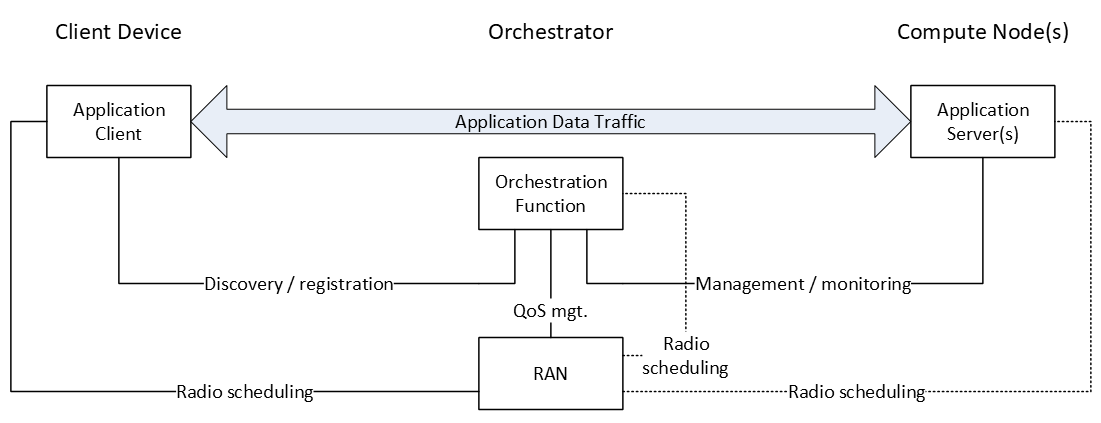} 
    \caption{High-level overview of the proposed ICC design.} \label{fig:ICC_architecture} 
    \vspace{-4mm}
\end{figure}

\subsection{5G MEC Overview}
The 3GPP EDGE architecture, defined in TS 23.558~\cite{3GPP_EDGE}, brings compute and application processing closer to end users. It adopts a model based on an operator-managed data network, conceptually resembling the over-the-top (OTT) model familiar from cloud computing. Edge Application Servers (EAS) reside behind a user plane function (UPF), which is located close to the base station, reducing latency and enabling localized service processing. An Edge Enabler Server (EES) and an Edge Configuration Server (ECS) coordinate discovery, instantiation, and session setup for the Edge Enabled Client (EEC) and the hosted applications, creating an environment that streamlines offloaded compute between the client and the operator's data network. Despite these advantages, this architecture has limitations:
\begin{itemize}[leftmargin=*]
    \item \textbf{Deployment Restriction:} It only supports EAS and EES deployments in operator-managed sites behind a UPF, which limits the proximity of compute to end users and rules out RAN-based and device-based servers.
    \item \textbf{Lack of RAN/Device Integration:} There is also no interface to orchestrate resources in the RAN or on the device side, hindering real-time, fine-grained decisions based on RAN conditions.
    \item \textbf{Static QoS Policies:} As no dedicated interfaces exist between compute entities and the RAN, the architecture cannot fully manage communication and computing latencies jointly. While it supports QoS-based session setup, the reliance on static or pre-configured policies prevents dynamic adjustment of user plane routes or computing resources in response to instantaneous demand or congestion, limiting the ability to seamlessly coordinate communication and computing.
\end{itemize}

\subsection{6G ICC Design}
To address the limitations of 5G MEC—where compute services do not benefit from close coupling with the cellular system—we propose an architecture for \textit{Integrated Communication and Computing} (ICC). ICC allows for a unified approach to compute offloading across multiple deployment scenarios, including OTT platforms, peer devices, the operator's RAN cloud, and MEC environments, with orchestration visibility to both computing and communication resources. In designing a 6G system that supports new QoS/QoE requirements for advanced services (e.g., joint communication and computing latency), the core principle of ICC is to integrate computing awareness across UEs, RAN, and the core network, while enforcing a robust QoE/SLA mechanism to guarantee the service needs.

Like the MEC architecture, an ICC framework depends on nodes with three roles, as shown in Fig.~\ref{fig:ICC_architecture}: the client device that hosts the user-facing application (generally analogous to the EEC in the MEC architecture), one or more computing nodes that offer compute services to the client (analogous to the EASs), and an orchestrator that coordinates the process (comprising functions similar to the EES and ECS). One node may perform several roles: A client device may perform its own orchestration of nearby compute resources, or an orchestrator may be hosted at a node that also provides resources as a computing node. ICC departs from the MEC model in allowing any role to be instantiated anywhere—in the network or at a device—and in providing the orchestrator with real-time visibility into both the performance of the computing nodes and the end-to-end performance of the service as a whole, along with dynamic influence over the communication QoS. This enhanced visibility and control allow the ICC model to adapt communication and compute functionality to one another in an ongoing way, according to the needs of the service.

There are two principal approaches that can enable ICC:
\begin{enumerate}[leftmargin=*]
    \item \textbf{Extending 5G MEC:} One way to realize ICC is by extending 5G MEC through additional network functions, instantiation of the node functions at devices as well as network nodes, enhanced user-plane capabilities for flexible routing of computing data, and exposure of compute and communication performance to the orchestrator via network and device interfaces.
    \item \textbf{New Data Plane:} Another approach is to introduce a new data plane that provides data processing, computing resource orchestration, data storage, and multi-hop processing for computing tasks, while allowing data delivery from point to point anywhere in the system.
\end{enumerate}
The ICC concept works with either model, requiring only some mechanism for communication between the involved nodes. By deploying ICC, operators can move computing closer to end users either at the network edge or on the device itself, enabling direct offloads and providing MNO-managed edge services. 
\label{sec:ICC}
\section{Theoretical Analysis}\label{sec:theory}
\begin{figure}[t] 
    \centering
    \includegraphics[width= 0.8\columnwidth]{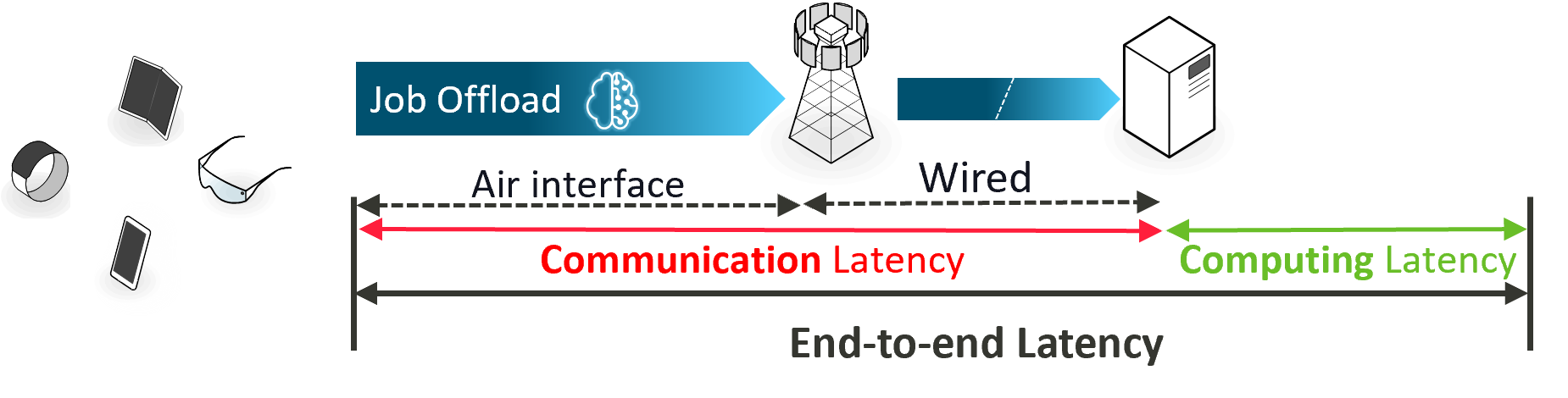} 
    \caption{6G ICC system model for theoretical analysis.} \label{fig:sys}
    \vspace{-3mm}
\end{figure}

\subsection{System Model}
We consider an ICC system, as illustrated in Fig.~\ref{fig:sys}, where users offload their computational jobs in an online manner and transmit the corresponding data to a computing node (potentially located at a RAN node, MEC node, or elsewhere) for processing. Specifically, offloading includes: (i) uplink wireless transmission of the job data from user equipment (UE) to a base station (BS), (ii) wireline transmission of the data from the BS to the computing node, and (iii) execution of the job by the computing node. Given that the downlink channel typically has greater capacity than the uplink, downlink latency (i.e., the output to UE) is negligible compared to the end-to-end latency budget. We therefore omit the downlink latency in this analysis. Each computational job must be completed within a prescribed latency budget.

Each offloaded job is assumed to have a uniform data size (for communication) and a fixed computational load at the computing node. The arrival of these jobs follows a Poisson process with rate~$\lambda$. The uplink service time on the air interface follows an exponential distribution with rate parameter $\mu_1$ (i.e., $\mu_1$ jobs per unit time can be transmitted from the UE to the BS). We denote the fixed wireline transmission delay between the BS and the computing node as $t^{\textrm{wireline}}$, determined by their physical distance. The computing procedure follows an exponential service time with rate parameter $\mu_2$ (i.e., $\mu_2$ jobs per unit time can be executed by the computing node).

To further compare the latency management effects between 5G MEC and ICC, we first introduce the concepts of disjoint and joint latency managements as follows:
\begin{itemize}[leftmargin=*]
    \item \textbf{Disjoint Latency Management:} Given a total latency budget, the system allocates a latency budget for each of the communication and computing procedures. Job execution is considered successful if both communication and computing latencies are within their respective budgets.
    \item \textbf{Joint Latency Management:} The system uses the entire latency budget for both communication and computing procedures. Job execution is considered successful if the end-to-end latency is smaller than the total latency budget.
\end{itemize}
We denote $T_{\textrm{E2E}}$ as the end-to-end latency for a computational job including communication and computing procedures, where the latency over air-interface is denoted by $T^{\textrm{UE-BS}}_{\textrm{comm}}$, the wireline latency is denoted by $T^{\textrm{wireline}}_{\textrm{comm}}$ and $T_{\textrm{comp}}$ is the computing latency, can be written as follows:
\begin{align}
    T_{\textrm{E2E}} = T^{\textrm{UE-BS}}_{\textrm{comm}} + T^{\textrm{wireline}}_{\textrm{comm}} + T_{\textrm{comp}} \label{eq:time_E2E}.
\end{align}
Given the total latency budget $b_{\textrm{total}}$, the goal is to ensure that each job can be executed within this latency budget, i.e., $T_{\textrm{E2E}} \leq b_{\textrm{total}}$, as formalized in the following definition:
\begin{definition}[Job Satisfaction]\label{def:job_satisfaction}
    A computational job is declared as a satisfied job if its end-to-end latency does not exceed the total latency budget, i.e., $T_{\textrm{E2E}}\leq b_{\textrm{total}}$.
\end{definition}
To quantize the highest rate of arriving jobs under which the system can meet the latency budget with a desired probability, we introduce the following key performance indicator (KPI).\footnote{We assume the system operates in steady state.}

\begin{definition}[Service Capacity]\label{def:service_capacity}
     Let \(\mathcal{E}(\lambda)\) be the event that a randomly chosen computational job (in steady state) is satisfied. Let \(\alpha\) be a target probability threshold (e.g., \(95\%\)). The \emph{service capacity} \(\lambda^*\) is defined as
    \begin{equation}
        \lambda^* \;=\; \sup \Bigl\{ \lambda : \mathbb{P}(\mathcal{E}(\lambda)) \;\ge\;\alpha \Bigr\}.
    \end{equation}
\end{definition}


\subsection{Theoretical Results}\label{sec:theoretical_result}
Given the wireline latency \(t^{\textrm{wireline}}\) and a total latency budget \(b_{\textrm{total}}\), let \(\mathcal{E}_{\textrm{Joint}}(\lambda, t^{\textrm{wireline}}, b_{\textrm{total}})\) be the event that a randomly chosen computational job (in steady state) is satisfied under joint latency management. Under disjoint latency management, we define \(\mathcal{E}_{\textrm{Disjoint}}(\lambda, t^{\textrm{wireline}}, b_{\textrm{total}}, b_{\textrm{comm}}, b_{\textrm{comp}})\) as the event that a randomly chosen computational job (in steady state) is satisfied, given that the latency budget \(b_{\textrm{total}}\) is split into \(b_{\textrm{comm}}\) for communication and \(b_{\textrm{comp}}\) for computing.
Building on~\eqref{eq:time_E2E} and the joint/disjoint latency management principles, for a given wireline latency \(t^{\textrm{wireline}}\) and total latency budget \(b_{\textrm{total}}\), the job satisfaction rate under joint latency management is given by 
\begin{align}
    &\mathbb{P}(\mathcal{E}_{\textrm{Joint}}(\lambda, t^{\textrm{wireline}}, b_{\textrm{total}})) = \mathbb{P}(T_{\textrm{E2E}} \leq b_{\textrm{total}}|T^{\textrm{wireline}} = t^{\textrm{wireline}})\nonumber\\
    &= \mathbb{P}(T^{\textrm{UE-BS}}_{\textrm{comm}}  + T_{\textrm{comp}} \leq b_{\textrm{total}} - t^{\textrm{wireline}}). \label{eq:cap_joint}
\end{align}
Meanwhile, under the disjoint latency management, the service capacity is given by 
\begin{align}
    &\mathbb{P}(\mathcal{E}_{\textrm{Disjoint}}(\lambda, t^{\textrm{wireline}}, b_{\textrm{total}}, b_{\textrm{comm}}, b_{\textrm{comp}}))\nonumber\\ 
    = &\mathbb{P}(T_{\textrm{E2E}} \leq b_{\textrm{total}}, T^{\textrm{UE-BS}}_{\textrm{comm}} + T^{\textrm{wireline}}_{\textrm{comm}}\leq b_{\textrm{comm}},  \nonumber\\
    &\qquad \qquad \qquad T_{\textrm{comp}} \leq b_{\textrm{comp}}|T^{\textrm{wireline}} = t^{\textrm{wireline}}) \nonumber\\
    = &\mathbb{P}(T^{\textrm{UE-BS}}_{\textrm{comm}} + T_{\textrm{comp}} \leq b_{\textrm{total}}- t^{\textrm{wireline}}, \label{eq:cap_disjoint} \\
     & \qquad \qquad T^{\textrm{UE-BS}}_{\textrm{comm}} \leq b_{\textrm{comm}} - t^{\textrm{wireline}}, T_{\textrm{comp}} \leq b_{\textrm{comp}}) \nonumber.
\end{align}

\begin{figure}[t] 
    \centering
    \includegraphics[width= 0.8\columnwidth]{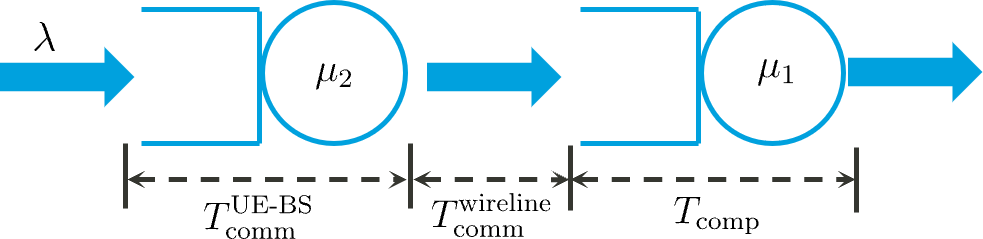} 
    \caption{Queueing modeling for 6G ICC system illustrated in Fig. \ref{fig:sys}.} \label{fig:queue}
    \vspace{-3mm}
\end{figure}

\textbf{Queueing Network Modeling:} To derive job satisfaction rates followed from Eq.~\eqref{eq:cap_joint} and \eqref{eq:cap_disjoint}, the joint probability distribution function (PDF) of $T^{\textrm{UE-BS}}_{\textrm{comm}}$ and $T_{\textrm{comp}}$ is required. As illustrated in Fig.~\ref{fig:queue}, the system can be modeled as a tandem queueing network comprising two service stages with rates \(\mu_1\) and \(\mu_2\). The first queue corresponds to the air interface communication procedure, and the second queue represents the computing procedure. In particular, jobs leaving the communication queue incur a constant wireline latency \(t^{\textrm{wireline}}\) before entering the computing queue. Lemma~\ref{lemma:independence} demonstrates the independence of sojourn times (the sum of waiting and service times) in the communication and computing queues.\footnote{The proof can be derived via Burke's theorem~\cite{burke1956output} and basic properties of the $M/M/1$ queue and Poisson. We omit the proof due to page limits.}


\begin{lemma}\label{lemma:independence}
    In steady state, under First-Come First-Served (FCFS) discipline, the computing queue is a $M/M/1$ queue. Moreover, the sojourn times experienced by any tagged job in the communication and computing queues are independent.
\end{lemma}

The sojourn time in an $M/M/1$ queue with Poisson arrival rate \(\lambda\) and service rate \(\mu\) follows an exponential distribution with parameter \(\mu - \lambda\)~\cite{stewart2009probability}. By Lemma~\ref{lemma:independence}, \(T_{\mathrm{comm}}^{\mathrm{UE\text{-}BS}}\) and \(T_{\mathrm{comp}}\) are therefore independent and  exponentially distributed with rate \(\mu_1 - \lambda\) and \(\mu_2 - \lambda\) respectively. Consequently, their joint PDF $f_{T_{\textrm{comm}}^{\textrm{UE-BS}},T_{\textrm{comp}}}$ is given by  
\begin{align}
    f_{T_{\textrm{comm}}^{\textrm{UE-BS}},T_{\textrm{comp}}}(t_{\textrm{comm}}^{\textrm{UE-BS}}, t_{\textrm{comp}})
    = f_{T_{\textrm{comm}}^{\textrm{UE-BS}}}(t_{\textrm{comm}}^{\textrm{UE-BS}}) \cdot f_{T_{\textrm{comp}}}(t_{\textrm{comp}})\\
    =  (\mu_1 - \lambda)e^{-(\mu_1 - \lambda)t_{\textrm{comm}}^{\textrm{UE-BS}}}\cdot(\mu_2 - \lambda)e^{-(\mu_2 - \lambda)t_{\textrm{comp}}} \label{eq:prob}
\end{align}
where \(f_{T_{\textrm{comm}}^{\textrm{UE-BS}}}\) and \(f_{T_{\textrm{comp}}}\) are the marginal exponential PDFs corresponding to rates \(\mu_1 - \lambda\) and \(\mu_2 - \lambda\) respectively.

\begin{figure}[t] 
    \centering
    \includegraphics[width= \columnwidth]{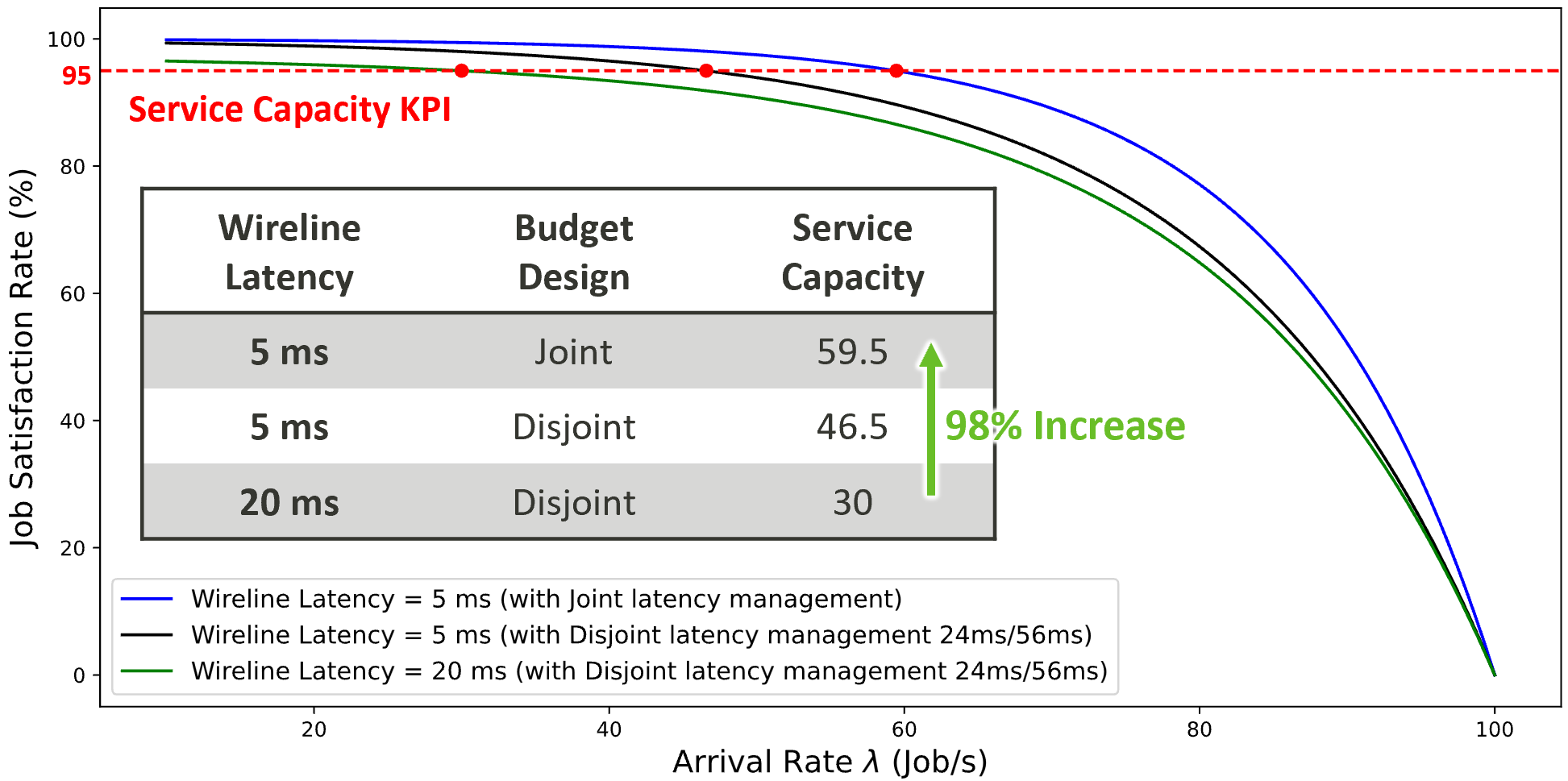} 
    \caption{Comparison of ICC and 5G MEC performance of job
    satisfication rate over different job arrival rates.} \label{fig:theory_capacity}
    \vspace{-3mm}
\end{figure}

Next, we consider a computational job service with an end-to-end latency budget of \(b_{\textrm{total}} = \text{80ms}\) running on an ICC system characterized by \(\mu_1 = \text{900}\) and \(\mu_2 = \text{100}\). We compare three schemes as follows: 
\begin{itemize}[leftmargin=*]
    \item Joint latency management with an RAN computing node (\(t^{\textrm{wireline}} = \text{5ms} \)), which flexibly utilizes the entire \(b_{\textrm{total}}\) for both communication and computing.
    \item  Disjoint latency management with an RAN computing node (\(t^{\textrm{wireline}} = \text{5ms} \)), which allocates \(b_{\textrm{comm}}= \text{24ms}\) for communication and \(b_{\textrm{comp}} =\text{56ms}\) for computing. 
    \item Disjoint latency management with a MEC computing node (\(t^{\textrm{wireline}} = \text{20ms} \)), which allocates \(b_{\textrm{comm}}= \text{24ms}\) for communication and \(b_{\textrm{comp}} =\text{56ms}\) for computing. 
\end{itemize}
Based on Eq.~\eqref{eq:cap_joint},~\eqref{eq:cap_disjoint} and~\eqref{eq:prob}, Figure~\ref{fig:theory_capacity} illustrates the resulting job satisfaction rates. The figure demonstrates that the service capacity ($\alpha = 95\%$ in this case) is increased up to $98\%$ under joint latency management, and by moving computing closer to the user.

\section{Simulation Analysis}\label{sec:sim}
\begin{figure*}
      \centering
      \includegraphics[width=0.85\textwidth]{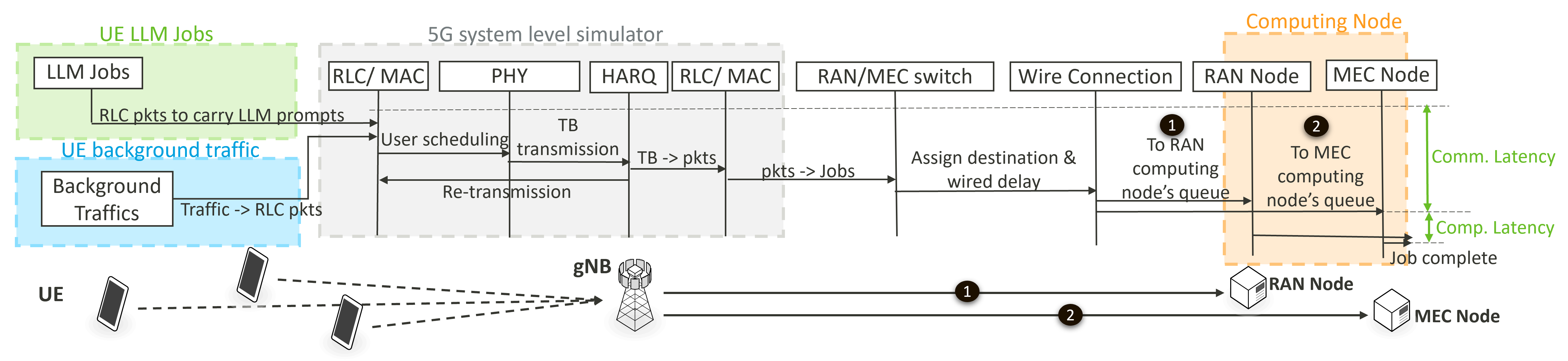}
      \caption{Overview of the simulation framework for performance evaluation of ICC compared to 5G MEC.}
      \label{fig:sim_framework}
      \vspace{-4mm}
\end{figure*}

In this section, we evaluate the performance of ICC in a simulated 5G environment and compare it against traditional 5G MEC. We consider a single gNodeB (gNB) serving multiple randomly deployed UEs. Each UE is equipped with an augmented reality (AR) glasses, which supports real-time language translation. Conversational texts are offloaded to a LLM running at a computing node. 
Each translation job is defined as $J = \{ N_{\textrm{input}}, N_{\textrm{output}}, C_{\textrm{LLM}}, M_{\textrm{LLM}}, b_{\textrm{total}} \}$, where $N_{\textrm{input}}$ and $N_{\textrm{output}}$ are the input and output prompt sizes, respectively; $C_{\textrm{LLM}}$ is the LLM's compute complexity; $M_{\textrm{LLM}}$ is the LLM's model size; $b_{\textrm{total}}$ is the job's end-to-end latency budget.

\subsection{Latency Modeling}
Following Section~\ref{sec:theory}, we consider the end-to-end latency $T_{\textrm{E2E}}$ as the timespan from the generation of input prompt to the completion of output prompt (i.e., the translation result).  Consequently, $T_{\textrm{E2E}}$ is determined as in Eq.~\ref{eq:time_E2E}, where $T^{\textrm{UE-BS}}_{\textrm{comm}}$ and $T^{\textrm{wireline}}_{\textrm{comm}}$ denote the uplink wireless and wired transmission latencies, respectively for transmitting the input prompts to the computing node, respectively, while $T_{\textrm{comp}}$ represents the latency incurred by the LLM inference stage. 

\noindent{\bf Communication Latency:} To measure $T^{\textrm{UE-BS}}_{\textrm{comm}}$, we've implemented a system level simulator (SLS) following the methodology in~\cite{GonzalezD2022}. In this simulator, the input prompts are first converted into RLC packets and transmitted to gNB over 5G physical layer, using certain channel realization and protocols. Therefore, for each packet, the wireless transmission latency through this simulator accounts for both transmission and queueing delays. Once the gNB receives the packets, they will be forwarded via a wired link to either the RAN or the MEC computing node. The wireline latency $T^{\textrm{wireline}}_{\textrm{comm}}$ for each packet between gNB and the computing node is modeled as a constant. 

\noindent{\bf Computing Latency:} The computing latency consists of two primary components: LLM inference latency and queueing delay of the job that stays at the computing node. LLM inference operations typically comprise two phases~\cite{llm_inference_guide}:
\begin{enumerate}[leftmargin=*]
  \item \textbf{Prefill Phase:} During the prefill phase, the model processes the entire batch of input tokens to create a key-value (KV) cache, which is later reused during the phase of output token generation to avoid redundant computations. Given that each LLM inference processes $N_{\textrm{input}}$ input tokens, the total computing complexity (in number of floating-point operations) for matrix multiplication can be approximated by $C_{\textrm{prefill}} = N_{\textrm{input}} \cdot C_{\textrm{LLM}}$ where $C_{\textrm{LLM}}$ is typically taken to be twice the number of model parameters. Due to the large size of modern LLMs, the model must also be loaded from memory, introducing non-negligible overhead given the large model size defined as $M_{\textrm{LLM}}$. Hence, the overall latency for the prefill phase is given by:
  \begin{equation}
  T_{\textrm{prefill}} = \max \Bigl\{ \frac{C_{\textrm{prefill}}}{G^{\textrm{comp}}_{\textrm{BW}}}, \frac{M_{\textrm{LLM}}}{G^{\textrm{memory}}_{\textrm{BW}}} \Bigr\}
  \end{equation}
  where $G^{\textrm{comp}}_{\textrm{BW}}$ is GPU's computational capacity (in FLOPS) and $G^{\textrm{memory}}_{\textrm{BW}}$ is the memory bandwidth (in bytes/s) for loading model parameters from GPU memory.
  \item \textbf{Output Token Generation Phase:} Because each output token requires the KV cache values of all the input tokens and preceding generated output tokens, the output tokens are produced one by one sequentially. Let $N_{\textrm{output}}$ be the number of output tokens. The overall latency for output token generation phase can be expressed as:
\begin{align}
T_{\textrm{tokengen}} = N_{\textrm{output}} \cdot \max \Bigl\{ \frac{C_{\textrm{LLM}}}{G^{\textrm{comp}}_{\textrm{BW}}}, \frac{M_{\textrm{LLM}}}{G^{\textrm{memory}}_{\textrm{BW}}}\Bigr\}.
\end{align}
\end{enumerate}
Therefore, the total LLM inference latency for a given job $J$ is $T_{\textrm{comp}} = T_{\textrm{prefill}} + T_{\textrm{tokengen}}$.

\subsection{Simulation Procedure}\label{sec:sim_procedure}
The complete simulation procedure is illustrated in Fig.~\ref{fig:sim_framework}. Each UE is randomly placed within the coverage area of gNB. Real-time translation jobs arrive at each UE following a Poisson distribution with arrival rate $\lambda$. Once a job $J$ is arrived, the UE generates uplink packets based on the input size of $N_{\textrm{input}}$ and transmits them to the gNB over the 5G air interface. The gNB forwards the job $J$ to either the RAN and MEC computing node once it receives all the relevant packets. The time between job forwarding and arrival at the computing node's queue is dependent on the wired latency. The job then waits at the computing node's queue until the computing node become available to process it. The time between job arrival at UE and job arrival at a computing node queue is measured as $T^{\textrm{UE-BS}}_{\textrm{comm}}$, the following time until the job completion is measured as $T_{\textrm{comp}}$. 

As described in Section~\ref{sec:ICC}, ICC can jointly manage communication and computing latencies, creating opportunities for integrated resource optimization. To leverage this potential, we implement a priority-based scheme with two key components:
\begin{itemize}[leftmargin=*]
  \item {\bf Job-aware Packet Prioritization:}  Because the job's characteristics are now transperant to the communication system, packets can be prioritized based on job requirements. When a real-time translation job arrives, its packets receive higher priority than background traffic packets.
\item {\bf Priority-Based Job Queueing:} The computing node can incorporate communication conditions into its queueing strategy by adopting a priority queue rather than following a simple FIFO order. Specifically, the priority of each arrival job is determined by its observed commmunication latency $T^{\textrm{UE-BS}}_{\textrm{comm}}$, its generation time $T^{\textrm{gen}}_{J}$, and its end-to-end latency budget $b_{\textrm{total}}$. In our simulation, we set the priority to the value of $T^{\textrm{gen}}_{J}$ + $b_{\textrm{total}}$ - $T^{\textrm{UE-BS}}_{\textrm{comm}}$. Any job expected to leave the computing node's queue after $T^{\textrm{gen}}_{J}$ + $b_{\textrm{total}}$ is dropped.
\end{itemize}

\subsection{Simulation Results}
\begin{figure}
      \centering
      \includegraphics[width=0.95\linewidth]{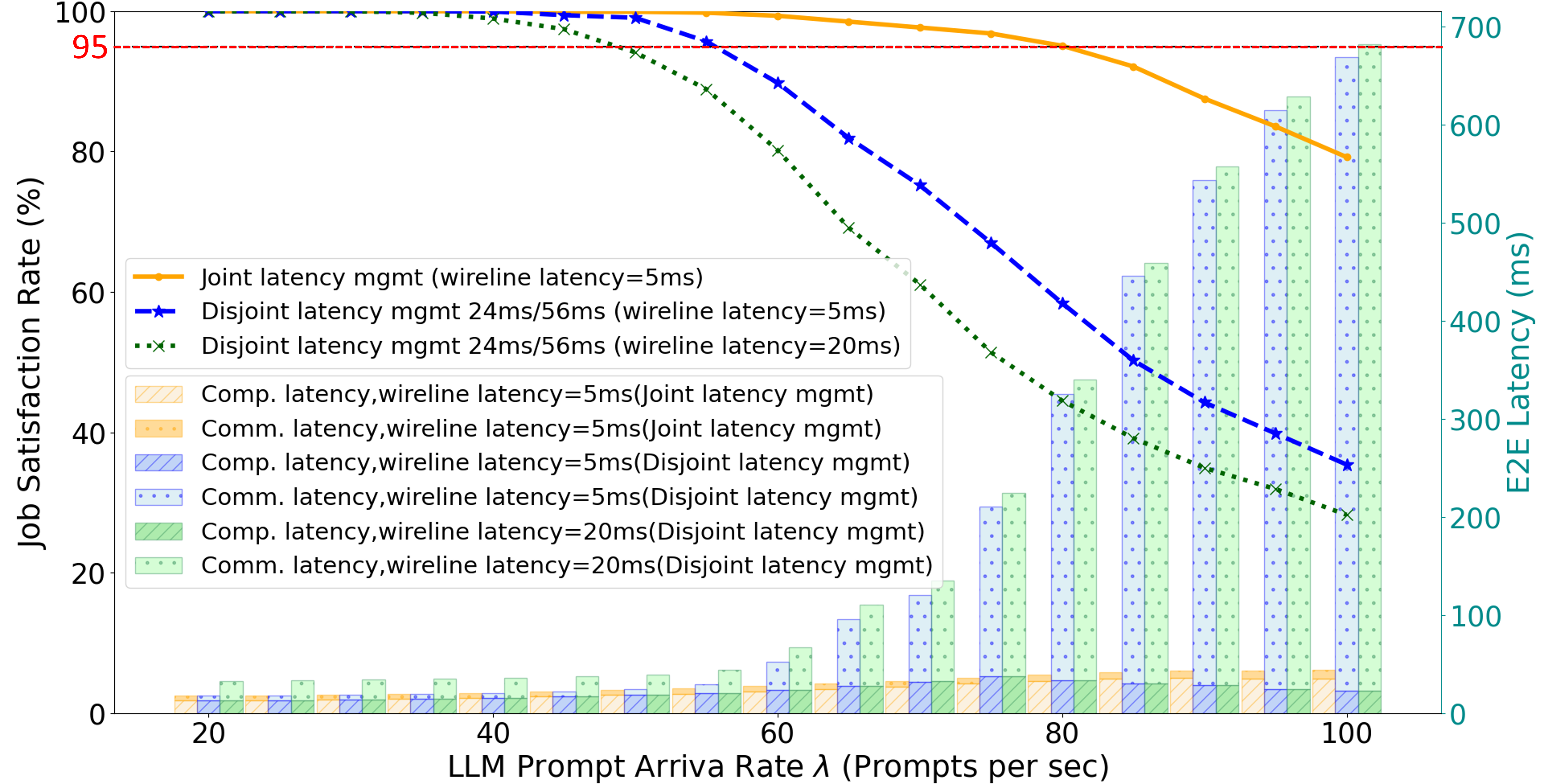}
      \caption{Comparison of ICC and 5G MEC performance of job satisfication rate over different prompt arrivals.}
      \label{fig:sim_result_arrivalrate}
      \vspace{-2mm}
\end{figure}

First, we compare the service capacity of ICC against 5G MEC under varying prompt arrival rates, as shown in Fig.~\ref{fig:sim_result_arrivalrate}. In this setup, each UE generates at a rate of 1 prompt/s and we scale the number of UEs to adjust the overall prompt arrival rate. We assume that each prompt consists of 15 input tokens, leading to 15 output tokens from LLM. The computing node is equipped with two GH200-NVL2 GPUs~\cite{GPU_SPEC_GH200}, and the rest of the parameters follow Table~\ref{tab:sim_setup}. We consider the LLM service with an end-to-end latency budget \(b_{\textrm{total}} = \text{80ms}\), comparing three schemes as outlined in Section~\ref{sec:theoretical_result}. As seen in Fig.~\ref{fig:sim_result_arrivalrate}, reducing the wireline latency from 20ms to 5ms (under the disjoint latency management) increases the service capacity ($\alpha = 95\%$) by $10\%$ that demonstrates the benefit of moving the computing node closer to the users. By incorporating joint latency management and maintaining wireline latency 5ms, the service capacity improves even further. Specifically, ICC supports the service capacity of up to $80$ prompts/s, whereas 5G MEC (disjoint latency management with 20ms of wireline latency) can only support 50 prompts/s, yielding a $60\%$ improvement. The bar plot in Fig.~\ref{fig:sim_result_arrivalrate} (Y-axis on the right hand side) shows the corresponding average computing and communication latencies across all the prompts. The communication latency climbs significantly as the prompt arrival rate increases, highlighting the advantages of ICC and the further gains achievable through the priority-based scheme as presented in Section~\ref{sec:sim_procedure}.

\begin{figure}[!t]
  \centering
  \includegraphics[width=0.95\linewidth]{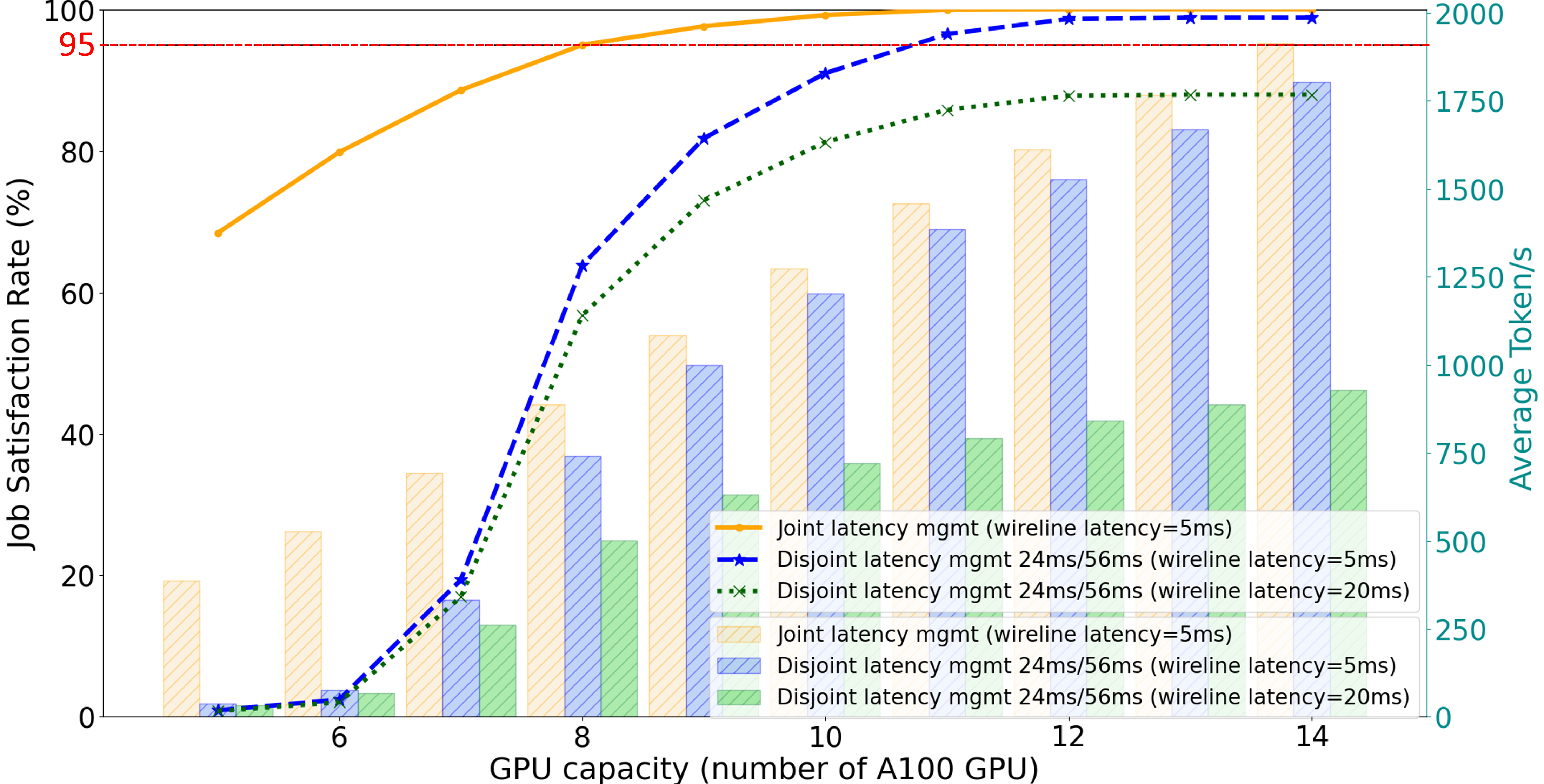}
  \caption{Job satisfication rate and average token per second comparison of ICC and 5G MEC over different computing node capacity.}
  \label{fig:sim_result_gpu_a}
  \vspace{-1mm}
\end{figure}

Next, Fig.~\ref{fig:sim_result_gpu_a} examines service capcity with different computing capacities at the computing node, where each scaled relative to a single Nvidia A100 GPU~\cite{GPU_SPEC_A100}. Here, the number of UEs is 60, and each generate prompts at 1 prompt/s, while other parameters remain as in Table~\ref{tab:sim_setup}. The curves show job satisfication rates. Disjoint latency management with a 20ms wireline latency fails to achieve a $95\%$ job satisfaction rate, even at high computing capacities. With wireline latency reduced to 5ms, the service capacity can reach the a $95\%$ job satisfaction rate with 11 A100 GPUs. This demonstrates the necessity of moving the computing node closer to users. In contrast, ICC attains a $95\%$ job satisfaction rate with only 8 A100 GPUs, yielding a $27\%$ saving in hardware costs. The bar plot in Fig.~\ref{fig:sim_result_gpu_a} shows the average tokens per second (total input and output tokens divided by end-to-end latency), revealing that lower wireline latency amplifies performance gains. Note that as GPU capacity increases, the performance disparity between joint and disjoint latency management diminishes. This indicates that in cloud computing environments, where computing resources are not a limiting factor, there is minimal incentive to integrate communication and computing system management. Conversely, in edge/RAN computing scenarios, ICC can more effectively leverage the constrained computational resources by jointly managing these systems, as evidenced by the performance improvements in the low GPU capacity region depicted in Fig.~\ref{fig:sim_result_gpu_a}.


\begin{table}[!t]
    \centering
    \begin{tabular}{| m{12 em} | m{8em} |}
      \hline
      {\bf Parameter} & {\bf Value} \\
      \hline
      Channel & Urban Macrocell \\
      \hline
      Carrier frequency & 3.7 GHz \\
      \hline
      Subcarrier Spacing & 60kHz \\
      \hline
      Bandwidth & 100 MHz \\
      \hline
      Background Traffic & 0.5 Mbps/UE \\
      \hline
      Job arrival rate $\lambda$ & 1/s/UE \\
      \hline
      LLM model & Llama 2 7B (FP16) \\
      \hline
      Input prompt size & 15 tokens  \\
      \hline
      Output prompt size & 15 tokens  \\
      \hline
    \end{tabular}
    \caption{Summary of key simulation settings}
    \label{tab:sim_setup}
    \vspace{-5mm}
\end{table}

\section{Conclusion}
\vspace{-2mm}
In this paper, we introduced the concept of ICC to address the stringent latency requirements of LLM-based GenAI services in future 6G networks. By enabling computing capabilities directly at RAN nodes, ICC unifies both communication and computing resources, thereby outperforming 5G MEC in terms of service capacity and end-to-end latency under relatively modest compute investment. Beyond our theoretical insights, simulation-based findings further highlight the advantages of priority-based scheduling for joint latency management, yielding additional performance gains. We anticipate that the ICC principles can be readily applied to other latency-sensitive applications poised to emerge in the 6G era, motivating further development of ICC. Beyond this work, an interesting research direction involves enhancing performance through system-wide job offloading, fully capitalizing on ICC's ability to holistically utilize the distributed computing resources across a cellular network.


\bibliographystyle{ieeetr}
{\footnotesize \bibliography{references.bib}}

\end{document}